\documentstyle[12pt]{article}

\thicklines                         
\def\a{\alpha}

\def\d{\delta}
\def\e{\epsilon}                
\def\f{\phi}                    
\def\g{\gamma}

\def\l{\lambda}
\def\m{\mu}
\def\n{\nu}
\def\o{\omega}
\def\p{\pi}                     
\def\th{\theta}                  
\def\r{\rho}                    
\def\s{\sigma}                  
\def\t{\tau}

\def\x{\xi}

\def\F{\Phi}

\def\O{\Omega}

\def\cc{{\cal C}}
\def\cd{{\cal D}}
\def\ce{{\cal E}}
\def\cf{{\cal F}}

\def\ch{{\cal H}}   

\def\ck{{\cal K}}
\def\cl{{\cal L}}

\def\cn{{\cal N}}
\def\co{{\cal O}}

\def\ct{{\cal T}}

\def\cbo{{\,\raise-.15ex\Sc [\,}}                       

\def\sl#1{\rlap{\hbox{$\mskip 1 mu /$}}#1}      

\def\svev#1{\left\langle #1\right\rangle}       

\def\dt#1{{\buildrel {\hbox{\large .}} \over {#1}}}     
\def\ddt#1{{\buildrel {\hbox{\LARGE .\kern-2pt.}} \over {#1}}}

\def\beqn#1{ \renewcommand{\theequation}{#1}
             \begin{eqnarray} }
\def\eeqn{ \renewcommand{\theequation}{\arabic{equation}}
           \end{eqnarray} }

\def\beqr#1{ \setcounter{equation}{#1}
             \begin{eqnarray} }

\def\eeqr{\end{eqnarray}}
\def\NON{\nonumber\\}
\def\beqrabc#1{ \setcounter{equation}{0}
                \renewcommand{\theequation}{#1\alph{equation}}
                \begin{eqnarray} }
\def\beqrn#1#2{ \setcounter{equation}{#2}
                \renewcommand{\theequation}{#1.\arabic{equation}}
                \begin{eqnarray} }

\def\rf{ref.~\cite}

\def\APH#1{Ann. Phys. {\bf #1}}
\def\CMP#1{Comm. Math. Phys. {\bf #1}}

\def\NPB#1{Nucl. Phys. {\bf B#1}}

\def\PLB#1{Phys. Lett. {\bf B#1}}
\def\PRD#1{Phys. Rev. {\bf D#1}}
\def\PR#1{Phys. Rev. {\bf #1}}
\def\PRL#1{Phys. Rev. Lett. {\bf #1}}

\def\PTP#1{Prog. Theor. Phys. {\bf #1}}
\def\MPL#1{Mod. Phys. Lett. {\bf A #1}}

\def\sstyle{\scriptstyle}

\def\rhs{\mbox{r.h.s.} }

\def\ie{\mbox{i.e.} }

\def\frac#1#2{ {\sstyle {#1\over #2} } }
\def\det#1{{\rm det}\left(#1\right)}

\def\Tr{{\rm Tr}\,}

\def\half{{1\over 2}}

\def\det{{\rm det\,}}

\def\ba{\bar{a}}
\def\dtd#1{{\buildrel \mbox{$\scriptscriptstyle \circ $} \over {#1}}}
\def\frg{ \mbox{ }^{\scriptscriptstyle (4)}g}
\def\trg{ \mbox{ }^{\scriptscriptstyle (3)}g}
\def\tmg{ \mbox{ }^{\scriptscriptstyle (t)}g}

\begin{document}
\hfill WIS--94/10--Feb--PH
\par
\begin{center}
\vspace{15mm}
{\large\bf Dynamical-Space Regular-Time Lattice\\
and Induced Gravity}\\[5mm]
{\it by}\\[5mm]
Yigal Shamir\\
Department of Physics\\
Weizmann Institute of Science\\
Rehovot 76100, ISRAEL\\[15mm]
{ABSTRACT}\\[2mm]
  \end{center}
\begin{quotation}

  It is proposed that gravity may arise in the low energy limit of
a model of matter fields defined on a special kind of a dynamical random
lattice. Time is discretized into regular intervals, whereas the discretization
of space is random and dynamical. A triangulation is associated to
each distribution of the spacetime points using the flat metric of the
embedding space. We introduce a diffeomorphism invariant, bilinear scalar
action, but no ``pure gravity'' action.

  Evidence for the existence of a non-trivial continuum limit is provided
by showing that the zero momentum scalar excitation has a finite energy in the
limit of vanishing lattice spacing. Assuming the existence of localized
low energy states which are described by a natural set of observables,
we show that an effective curved metric will be
induced dynamically. The components of the metric tensor are
identified with quasi-local averages of certain microscopic properties of the
quantum spacetime. The Planck scale is identified with the highest mass scale
of the matter sector.

\end{quotation}

\newpage
\thispagestyle{empty}
\mbox{  }
\setcounter{page}{1}

\newpage
\noindent {\bf 1.~~INTRODUCTION}
\vspace{3ex}

  One of the fundamental problems in high energy physics is the
construction of a consistent quantum theory of gravity that will reduce
at large distances to  General  Relativity.
At the same time, such a quantum theory is expected to be
free of the diseases that arise when a straightforward
quantization of the classical theory is attempted.

  A natural approach to the definition of Quantum Gravity~[1-4]
is to consider models which are based on some
discretization of spacetime (see \rf{dtr} for a recent review).
Over the last  several years, important progress
has been made in two dimensions where, using the Dynamical Triangulation
(DT) approach~\cite{dt,dtr}, it was possible to prove the existence of a
continuum limit which reproduces known results on 2D gravity~\cite{2dg}.

  The basic idea of the DT approach, is to approximate $d$-dimensional
Riemannian manifolds by triangulations made of
equilateral $d$-simplices. The euclidean partition function is defined
as a sum over all such $d$-dimensional triangulations
weighted by some probability measure. Typically, the probability measure
is written as the exponential of (minus) an action. The sum over
triangulations is sometimes restricted to a fixed topology. We comment,
however, that a complete topological classification in four dimensions is
still an open problem.

  The advantage of the DT approach is that general coordinate invariance
is automatically satisfied, because the triangulations are defined without
any reference to a coordinate system, and a unique discrete Rimannian
structure can be assigned to each triangulation. Initially it was hoped that
the successes of the DT approach in two dimensions could be repeated in
four dimensions~\cite{4dt}. But gradually it was realized that
there are great difficulties with the DT approach in four dimensions,
which are related to the rapid growth in the number of triangulations as a
function of $N_4$, the number of four-simplices.

  In fact, recent
results~\cite{ckr} suggest that the number of distinct four dimensional
triangulations with the topology of a four-sphere is not
bounded by an exponential function of $N_4$. This implies that the grand
canonical partition function which involves a sum over $N_4$, does not
exists for conventional probability measures even if the sum is restricted
to triangulations with the topology of a four-sphere.
The inclusion of an arbitrarily large bare cosmological constant in
the action is not sufficient to compensate
for the rapid growth in the number of triangulations. Another result which
indicates that the space of four-triangulation is too big, is the
algorithmic unrecognizability of four-triangulations discussed in~\cite{nb}.

  The program of defining four dimensional Quantum Gravity
is therefore still struggling with the basic issue of
finding a consistent measure, and proving the existence of
some critical point which can be identified with a sensible continuum limit.

  If we ignore cosmological issues such as the expansion of the universe
for the time being, the observed ground state of the universe is, to
an extremely good accuracy, flat space. Moreover, the observed curved metric
is essentially classical, and one cannot detect any effect which can be
attributed to quantum fluctuations of the metric.

  These simple observational facts should provide a guideline in the search
of a consistent measure for Quantum Gravity. If the lack of progress
with conventional measures is due to the fact that the space
of four-triangulations is so huge, then what is needed is to
truncate that huge space into a much smaller one.
One would like a measure that enhances the contribution of smooth
triangulations over the contribution of irregular ones. Moreover,
to account for the classical character of the observed metric,
the dominant configurations should resemble only a {\it single} smooth
manifold, which would then be identified with the ground state.

  Suppose that a  measure with these good properties
can be found. How unique would such a measure be?
{}From our experience with quantum field theories on a regular lattice, we
are used to the concept of universality. Namely, many lattice actions which
differ in details have the same continuum limit, and the continuum limit
can be characterize by a small number of renormalized parameters such as
coupling constants and masses.

  But logically there could be other possibilities. Since the space of all four
dimensional triangulations is so huge, the case could be that many
different ``good'' measures can be defined on it. Each ``good''
measure would correspond to a different truncation of the space of
four-triangulations, and would define a different, but consistent, theory of
Quantum  Gravity through its continuum limit. In particular, each measure
would in general give rise to a different smooth manifold as the ground state.

  In fact, we cannot exclude the possibility that the space of all consistent
measures  may be as big as the space of smooth  manifolds which are candidates
for the ground state. If this is true, then the ground state cannot be
considered as a prediction of Quantum Gravity. Rather, under these
circumstances, the desired ground state would have to be {\it prescribed},
and would have to be supplied as an additional information in some way,
when selecting the appropriate measure for a phenomenologically viable model
of Quantum Gravity.

  There is a simple way to enhance the contribution of triangulations
that resemble a prescribed smooth four-manifold which we henceforth
denote as the target manifold. One can  allow only
for those triangulations which can be {\it embedded in the target
manifold}. This can be realized by demanding that all the vertices of
the triangulation should belong to the target manifold, and that
linking should be performed using some prescription which makes use of the
geodesic distance between points in the target manifold.
The partition function's measure is then defined as an integral over
the positions of all vertices in the target manifold.

  As long as the action is diffeomorphism invariant, in the sense that it
depends only on the connectivity of the resulting
triangulations (but not on the distances between vertices in the target
manifold) the partition function will define an effective measure on the
space of triangulations and, hence, {\it some} theory of gravity.
Apart from restricting the sum to a single topology, the virtue
of this prescription is that it assigns to each admissible triangulation an
entropy factor which significantly enhances the contribution of regular
triangulations. By this we mean that, as in classical statistical mechanics,
one can define thermodynamical fields such as the local density of points,
and on the dominant triangulations these fields will fluctuate very little
around their mean values throughout the entire target space.

  Moreover, if the target manifold is a symmetric space, then there is a
good chance that the ground state of the quantum theory will reproduce
that same space. Consider for example flat space as the target manifold.
A partition function constructed alone the lines described
above will be invariant under continuous {\it global} translations. If this
global symmetry remains unbroken (or if it is broken spontaneously leaving
a sufficiently large unbroken discrete
subgroup) the ground state found by low energy observers will be flat space.

  In this paper we consider a realization of the above ideas, taking
the target manifold to be flat space. (To define the finite volume partition
function we compactify flat space to a flat four-torus). The measure of the
euclidean partition
function is an integral over the positions of a fixed number of points
in the four-torus, where in the continuum limit the density of points should
tend to infinity. To each distribution of points we associate a triangulation
using a prescription which involves the flat norm on the four-torus.

  The plan of this paper is as follows. In sect.~2 we define the model and
establish some of its basic properties. The first guideline for the definition
of the model has already been discussed above. This is the requirement that
{\it the discretized spacetime structure is embedded in a fixed target manifold
which we take to be a flat four-torus}.

  Our second requirement is that {\it any matter action that we introduce
should depend only on the connectivity of the spacetime triangulations}.
Moreover, we {\it do not} introduce any ``pure
gravity'' action. The reason is that our spacetime measure already induces
an {\it effective} pure gravity action.

  The last feature is that {\it we treat time in a different way from space}.
Time is discretized into regular intervals, and only the discretization of
space is random and dynamical. We denote the resulting framework as
Dynamical-Space Regular-Time lattice (or DSRT lattice for short).
The special role of time allows us to
prove the existence of a euclidean transfer martix and, hence, to prove
unitarity of the time evolution in the continuum limit.
The reasons for the uneven treatment of space
and time will be clarified later. In short, we will argue that if one expects
to recover General Relativity at sufficiently low energies, then the special
role of time is not only a sufficient condition, but actually a
{\it necessary} condition for consistency of the continuum limit.

  In this paper we discuss only the simplest possible matter action.
This action describes a scalar field without self-interaction.
By deriving the Ward identity that corresponds to the broken shift symmetry
of the scalar action, we prove that  the zero momentum scalar excitation has
a finite energy in the limit of vanishingly small lattice spacing.
This provides a first indication that the model has a non-trivial continuum
limit. Strictly speaking, based on this result alone one cannot rule out the
possibility that the continuum limit might describe a system with a finite
number of states. But we believe that similar techniques, together with
variational estimates, can be used to show that the zero momentum state is
the end point of a continuous spectrum.

  In sect.~3 we investigate the feasibility that the continuum limit of the
DSRT model contains the familiar gravitational interaction.
A preliminary requisite is the construction of an appropriate set of
observables. There are well known difficulties associated with the definition
of local observables in gravitational theories, which we will not repeat here.

  In our approach, the embedding in the target manifold provides an additional
structure, which can be used in the definition of local observables.
The quantities which parametrize multi-local observables are taken to be
the coordinates of the $n$ points in the target manifold.
The classical continuum limit of the matter action has a natural
interpretation for such observables, and it allows us to identify the (inverse)
curved metric tensor with  {\it thermodynamical} fields that describe
quasi-local averages of certain  microscopic properties of spacetime.

  Our definition of observables and the ensuing identification of the curved
metric tensor with thermodynamical fields has important advantages, which are
discussed in subsect.~3.2. The price is that we introduce dependence on a
{\it coordinate system}. We argue that, by introducing an integration
over a small subset of the diffeomorphism group, one recovers general
coordinate invariance of the equation of motion in the continuum limit.

  The rest of sect.~3 is devoted to a discussion of the  resulting
dynamics of the gravitation field. In particular, we argue that the Planck
scale should be identified with the largest mass scale of the matter sector.
Other issues that we address include the nature of the
effective theory above the Planck scale, and recovery of Lorentz
covariance in the low energy limit.

  Some concluding remarks are given in sect.~4.
In particular, we offer several research directions to test key
features of the DSRT model.


\vspace{5ex}
\noindent {\bf 2.~~THE DYNAMICAL-SPACE REGULAR-TIME LATTICE}
\vspace{3ex}

  The formulation of quantum field theories on a Dynamical-Space Regular-Time
(DSRT) lattice is developed in this section. The random lattice points are
taken from a fixed underlying flat space compactified to a four torus.
Time is discretized into regular intervals, whereas the discretization
of space is random and dynamical. We assume a diffeomorphism invariant matter
action. In this paper we limit ourselves to the simplest case where the
matter action is bilinear in a set of scalar fields,

  The advantage of the DSRT lattice compared to the alternative approach
in which space and time are treated on equal footing, in that one can
establish rigorously certain fundamental
properties which are expected to hold in a consistent quantum field theory.
The special role of time allows us, using  standard
methods~\cite{tm},  to define a  Hilbert space of observables and to construct
a bounded, positive definite transfer matrix. This guarantees unitarity of the
time evolution in the continuum limit. The price paid is that restoration of
full Lorentz covariance is not automatic.

  We discuss the properties of the
ground state. A simple entropy argument provides evidence that the ground
state should be homogeneous when probed on a sufficiently large distance
scales, and that the ground state found by a low energy observer is flat space.
We next discuss the properties of the zero momentum scalar excitation.
We prove that this is a stable excitation, whose energy is exactly equal to
the mass parameter in the scalar action. This provides a first indication
for the existence of a non-trivial continuum limit.
Finally, we  show that spacetime fluctuations  induce an attractive interaction
among matter fields.

\vspace{2ex}
\noindent {\bf 2.1~~The partition function}
\vspace{1ex}

  In order to avoid excessive notation we define here the partition function
for a  DSRT lattice model containing a single scalar field.
(The generalization to several scalar fields is trivial). We will assume the
absence of scalar self-interactions.

    The canonical partition function depends on the following parameters:
$N$ is the total number of lattice points on every time slice and $L^3$ is the
space volume which we take to be a large cube. The discrete time coordinate $t$
takes the values $t=-N_0,-N_0+1,\ldots,N_0$, where the time slices  $t=\pm N_0$
are identified. Periodic boundary conditions are assumed in the space
directions too. The time interval between two neighbouring time slices is
$a_0$. Finally, $M$ is the mass of the scalar field. We define the mean
distance between space points through the relation
$$
   \ba=L/N^{1\over 3}\,,
\eqno(2.1)
$$
\ie  $1/\ba^3$ is the average spatial density of lattice points.

  The partition function is
\beqn{2.2}
  Z  & = & Z(N_0,a_0,N,L,M) \NON
     & = & \int_{n,t} \cd y  \int_{n,t} \cd \F \, e^{-S}\,.
\eeqn
Our notation conventions will be as follows. The subscrips $n,m=1,\ldots,N$
label the space points on every time slice, whereas the superscripts
$i,j,k,l=1,2,3$ denote space coordinates.
Thus, $y_{n,t}^i$ is the $i$-th space coordinates of the $n$-th point on the
time slice $t$, and  $\F_{n,t}$ is the value of the real scalar field residing
at this point. In order to avoid cumbersome expressions we use the shorthands
$$
  \int_{n,t} \cd y \equiv   \prod_{t=-N_0}^{N_0-1} \int_n \cd y_t \,,
$$
$$
  \int_n \cd y_t \equiv
    L^{-3N}  \prod_{n=1}^N \int_{-L/2}^{L/2} d^3 y_{n,t} \,,
$$
and
$$
  \int_{n,t} \cd \F \equiv   \prod_{t=-N_0}^{N_0-1} \int_n \cd \F_t \,,
$$
$$
  \int_n \cd \F_t \equiv
  \left( { a_0 \ba^3 M^2 \over 2\p} \right)^{N \over 2}
  \prod_{n=1}^N \int_{-\infty}^{\infty} d\F_{n,t} \,.
$$
The spacetime measure has been normalized to unity $\int_{n,t} \cd y = 1$.
The choice of normalization for the scalar measure is a matter of convenience.
Using the action~(2.5) below, one has $0<Z<1$.

  For every time slice (\ie for every copy of space) the measure describes an
integration over the locations of $N$ points in a flat three-torus.
The three-torus is the cube $L^3$ with opposite faces identified, and
it is equipped with the flat norm inherited from $R^{(3)}$.
The location of each point is defined by its flat coordinates $y^i_{n,t}$ and
the distance between any pair of points in the flat three-torus norm can
easily be determined from these coordinates.

  Because of the special role of time, the triangulations we build are not
made of four-simplices. Instead, on every time slice we build a triangulation
made of tetrahedra, whereas the temporal links provide a one-to-one mapping
between the points on two neighbouring time slices.

  In more detail, for a given set of space points
$\{y_{1,t}^i,\ldots,y_{N,t}^i\}$ a triangulation is constructed on every time
slice using the ``canonical'' prescription described
for example in ref.~\cite{col}. According to this prescription, four space
points are considered the vertices of a tetrahedron, and every pair of them is
joined by a link, provided  there are no other points inside their
circumscribed sphere. This prescription assigns a unique triangulation
to a given set of space points, except for degenerate cases that have zero
measure.

  In addition, every point on the time slice $t$ is connected to a single
point on each of the two neighbouring time slices $t\pm 1$. The linking between
the time slices $t$ and $t+1$ is described by a one-to-one mapping
$\t\!: (n,t) \to (m,t+1)$, where $\t((n,t)) = (m,t+1)$ if and only if the two
points $y_{n,t}$ and $y_{m,t+1}$ are connected by a link. The linking is
determined by requiring that the mapping $\t$ minimize the average
displacement squared
$$
  N^{-1} \sum_{n=1}^N \left| y_{\t(n,t)} - y_{n,t} \right|^2 \,.
\eqno(2.3)
$$
On the \rhs of eq.~(2.3) distances are measured in the flat three-torus norm,
and we have made use of the natural identification between the time slices
$t$ and $t+1$.

  Successive applications of the mapping $\t$ define a ``world line'' for each
space point. We will describe the (imaginary) time evolution of a given point
by
$$
  Y_{n,t}^i =  \left. y_{m,t}^i \right|_{(m,t)= \t^{t+N_0}(n,-N_0) } \,.
\eqno(2.4)
$$
We denote by $\F(Y_{n,t})$ the value of the scalar
field at the point $Y_{n,t}$. Hence,  $\F(Y_{n,t})=\F_{m,t}$
if $\t^{t+N_0}(n,-N_0) = (m,t)$. Notice that, in spite of the use of periodic
boundary conditions, $\t^{2N_0}(n,-N_0)$ will in general be different from
$(n,-N_0)$.

  While the prescription~(2.3) is clearly intended to prevent the occurrence
of very big displacements in the continuum limit, this prescription is
non-local
in space, in the sense that it relies on information about the entire
distribution of points on the two neighbouring time slices. By contrast,
the triangulation prescription on every time slice is strictly local.

  The non-locality in space of the time linking prescription is in fact
unavoidable, if we want to prevent the occurence of arbitrarily big
displacements. It arises because the mapping between the points of two time
slices is one-to-one. To see this, imagine that we take a pair of points which
are connected by a time link, and we move one of them to a new location which
is very far away. In order to prevent the occurence of macroscopically big
displacements in the new mapping, relinking has to take place over a
connected region which covers both the old location and the new one.

  Whether or not the non-locality of the time linking prescription
leaves a detectable trace in the
continuum limit, is a question that will have to be investigated in the future.
We insist on this prescription because, as argued in subsect.~3.1 below,
consistency of our identification of the dynamical curved metric tensor
requires that the fluctuations of world lines should obey a controlled bound.

  Having completed the construction of the {\it spacetime} triangulation that
corresponds to given sets of space points on every time slice, we can now
define the matter action. The scalar action is given by
$$
   S= a_0 \sum_{t=-N_0}^{N_0-1} \cl_t \,,
\eqno(2.5)
$$
where, suppressing the $t$-dependence,
$$
  \cl = {\ba^3\over 2} \sum_n \dtd\F_n^2 +
      {\ba\over 2} \sum_{\langle mn \rangle} (\F_m-\F_n)^2 +
      {M^2 \ba^3 \over 2} \sum_n \F_n^2 \,.
\eqno(2.6)
$$
In eq.~(2.6) the second term is a sum over the spatial links of the
triangulation, and the third term is a sum over over its vertices. The first
term is a sum over the temporal links, and it is the only one which
depends on a knowledge of the world line $Y_{n,t}$. For the coordinates of the
space points, we define the discrete time derivative as
$$
  \dtd{Y}^i_{n,t} = a_0^{-1} ( Y^i_{n,t+1} - Y^i_{n,t} ) \,,
\eqno(2.7)
$$
and for the scalar field we let
$$
  \dtd\F_n = a_0^{-1} \left( \F(Y_{n,t+1}) - \F(Y_{n,t}) \right) \,.
\eqno(2.8)
$$
Notice that, while each triangulation is constructed using the flat norm of
the target space, the matter action is diffeomorphism invariant. Namely, it
depends only on the connectivity of the triangulations, but
not on the metric properties of their embedding in the target space.

\newpage
\vspace{2ex}
\noindent {\bf 2.2~~The Hilbert space of observables and the transfer matrix}
\vspace{1ex}

  In the previous subsection we defined the partition function of scalar matter
on a DSRT lattice. Starting from the partition function $Z$ we now want to
reconstruct a {\it Hilbert space} of {\it observables} and a {\it bounded,
symmetric and positive definite transfer matrix}.

  Our definition of local observables relies on the embedding of the discrete
spacetime structure in $L^3 \times Z^{(2N_0)}$. The time $t$ observables are
defined as
$$
  \co(y^i_{n,t},\F_{n,t};f;\e,x^i) = f(\F_t(\e,x^i)) \,,
\eqno(2.9)
$$
where $f(\F)$ is a real polynomial and
$$
  \F_t(\e,x^i) = {3\ba^3\over 4\p\e^3} \sum_{n\in\cd(x;\e)} \F_{n,t} \,.
\eqno(2.10)
$$
In eq.~(2.10), $\cd(x;\e)$ is the ball of radius $\e$ centered at
$x^i\in R^{(3)}$, and $n\in\cd(x;\e)$ is a shorthand for summation over all
the time $t$ space points that belong to $\cd(x;\e)$.

  The above equations define observables which are local in both space and
time. In a similar way one can define multi-local observables, as well as
observables that involve the discrete time derivative of the scalar
field~(2.8). One can also consider observables that depend explicitly on
$y^i_{n,t}$.

  We define the  Hilbert space of observables to be the linear space of all
observables which depend only on dynamical variables with $0\le t\le N_0$.
In this Hilbert space we define a real inner product by
$$
  \Big(\!\!\Big( \co \Big| \co' \Big)\!\!\Big) \equiv
     \int_{n,t} \cd y  \int_{n,t} \cd \F \, e^{-S} (\Theta\co)\co'\,,
\eqno(2.11)
$$
where
$$
  \Theta\co(y_{n,t},\F_{n,t};f;\e_1,x_1;\e_2,x_2;\ldots) =
   \co(y_{n,-t},\F_{n,-t};f;\e_1,x_1;\e_2,x_2;\ldots) \,.
\eqno(2.12)
$$
In order to verify that the induced norm is strictly positive we have to
demonstrate reflection positivity around the $t=0$ plane. Namely, we have to
prove that the \rhs of eq.~(2.11) is strictly positive for $\co'=\co$. This
property will be established below.

  We next turn to the construction of the transfer matrix.
Let $\hat\ch$ be the  Time Zero Hilbert space, defined as the linear space of
all time zero observables
$$
\co(y_n,\F_n;f;\e_1,x_1;\e_2,x_2;\ldots)
$$
where now
$$
f(\F(\e_1,x_1),\F(\e_2,x_2),\ldots)
$$
is a square integrable function
with respect to the real inner product
$$
  \Big(\!\!\Big( \co \Big| \co' \Big)\!\!\Big)_0 =
   \int_n \cd y \int_n \cd \F\,
   \co(y_n,\F_n)\,\co'(y_n,\F_n)\,.
\eqno(2.13)
$$

 The transfer martix is a mapping  of $\hat\ch$ into
itself. For a given element of $\hat\ch$, we define
$$
  \left(\ct\co\right) (y'_m,\F'_m) =  \int_n \cd y \int_n \cd \F\,
  \ck(y'_m,\F'_m;y_n,\F_n)\,\co(y_n,\F_n)\,.
\eqno(2.14)
$$
The integral kernel is given by
$$
  \ck(y'_m,\F'_m;y_n,\F_n) = \ck_1(y'_m,\F'_m)\,\ck_2(y'_m,\F'_m;y_n,\F_n)\,
                        \ck_1(y_n,\F_n)\,,
\eqno(2.15)
$$
where
$$
  \ck_1(y_n,\F_n) = \exp\left\{ -{\ba a_0\over 4} \sum_{\langle mn \rangle}
          (\F_m-\F_n)^2 - {M^2 \ba^3 a_0 \over 4} \sum_n \F_n^2 \right\}\,,
\eqno(2.16)
$$
and
$$
  \ck_2(y'_m,\F'_m;y_n,\F_n) = \exp\left\{ -{\ba^3\over 2 a_0} \sum_{m,n}
          \left( \F'_m - \F_n \right)^2 \d_{n,\t(m)} \right\}\,.
\eqno(2.17)
$$
In eq.~(2.17) the mapping $\t$ between the primed and unprimed space points is
defined analogous to eq.~(2.3). We have explicitly denoted the dependence of
the
observable $\co$ on the dynamical variables. Notice that the complete set of
space points has to be included in the definition of the observable, otherwise
the mapping $\t$ which enters the integral kernel $\ck_2$ is undefined. This is
a manifestation of the non-locality in space of the time linking prescription.

   The transfer matrix is a manifestly bounded operator. It is also symmetric,
because the minimization prescription~(2.3) which defines the mapping $\t$ is
symmetric with respect to the two time slices. However, the integral kernel
$\ck_2$ is symmetric only up to a permutation of one of the indices $m$ or $n$.
Since this permutation depends on the coordinates of the space points, we are
unable to ``take the square root of $\ck_2$'', and so we are unable to prove
the positivity of the one step transfer matrix.

  Instead, we will now show that the
{\it two steps} transfer matrix $\ct^2$ is strictly positive. The same
reasoning will also prove reflection positivity around the $t=0$ hyperplane.
In order to prove the positivity of $\ct^2$, it is sufficient to prove
$$
  \Big(\!\!\Big( \co \Big| \ct^2\co \Big)\!\!\Big)_0 > 0 \,,
\eqno(2.18)
$$
for all non-zero operators $\co\in\hat\ch$. The validity of inequality~(2.18)
follows trivially by using $\ct^\perp = \ct$. For the reader who is worried
that this compact notation could be hiding some subtlety, we now explain in
detail how the positivity of $\ct^2$
can be established directly from its integral representation.

  Thinking of the variables $(y_n,\F_n)$ and $(y'_m,\F'_m)$ as living on the
Time Zero and Time One slices respectively, let us introduce a third set of
variables $(y''_p,\F''_p)$ that lives on the Time Two slice. $\ct^2$ is
represented by integrating over both $(y_n,\F_n)$ and $(y'_m,\F'_m)$ using
the integral kernel. Now, for the action of $\ct$ between the  Time Zero and
Time One slices we simply use eq.~(2.17) for
the integral kernel $\ck_2(y'_m,\F'_m;y_n,\F_n)$.
This integral kernel contains a factor $\d_{n,\t(m)}$. For the action of
$\ct$ between the  Time One and Time Two slices we have
$\ck_2(y''_p,\F''_p;y'_m,\F'_m)$ but with a factor $\d_{m,\bar\t(p)}$.
We distinguish between the {\it permutations} $\t$ and $\bar\t$.
Notice that $m=\bar\t(p)$ is the mapping {\it from} the Time Two slice {\it to}
the Time One slice, while $n=\t(m)$ is the mapping {\it from} the Time One
{\it to} the Time Zero slice. Rewriting
$\d_{m,\bar\t(p)}=\d_{p,\bar\t^{-1}(m)}$,  we observe that $p=\bar\t^{-1}(m)$
is the mapping from the Time One to the Time Two slice. The permutation
$\t=\t(y_n,y'_m)$ is determined by the coordinates of the Time Zero and Time
One points. The permutation $\bar\t^{-1}=\bar\t^{-1}(y''_p,y'_m)$ is
determined by the coordinates of the Time Two  and Time One points. We thus
see that $\t$ and $\bar\t^{-1}$ are {\it the same} function of their ordered
set of arguments. As a result, the integral representation of
$\big(\!\big( \co \big| \ct^2\co \big)\!\big)_0$ is equal to the integral
representation of  $\big(\!\big( \co \big| \ct^\perp\ct\co \big)\!\big)_0$.
This proves explicitly both the symmetry of the operator $\ct$ and the
positivity of $\ct^2$.

  As expected, the partition function is $Z=\Tr \ct^{2N_0}$, and in the
continuous time limit we can defined the hamiltonian through
$H=\lim_{a_0\to 0} {\bf :}H(a_0){\bf :}$ where
$$
  H(a_0) = - 1/(2a_0) \log \ct^2\,.
\eqno(2.19)
$$
The normal ordering symbol stands for a subtraction of the bare vacuum
energy density.

\vspace{2ex}
\noindent {\bf 2.3~~The ground state}
\vspace{1ex}

  We now proceed to discuss the properties of ground state. Let us first
examine the global spacetime symmetries of the DSRT lattice.
Thanks to the choice of periodic boundary conditions, the partition function
$Z$ is manifestly invariant under discrete time and continuous space
translations. In addition, the finite volume
partition function is invariant under a discrete subgroup of space rotations.
(In the infinite volume limit full rotation symmetry is recovered).
If none of these symmetries are spontaneously broken, then the ground state
will be flat space.

  In order to tell whether or not translation invariance is broken in some
region of the phase diagram of the model, one has to carry out a detailed
dynamical calculation. But at this stage we do not really care whether the
ground state is {\it exactly} translationally invariant. What we care about is
whether the ground state {\it as probed by a low energy measurement} cannot
be distinguished from a homogeneous, continuous flat space. For example,
an interesting possibility is that solidification may take place~\cite{epr},
resulting in a regular crystal-like structure of the ground state.
But if the cell of
that crystal is of Planck size, then low energy excitations will be blind to
the existence of that microscopic structure, and the ground state of the
low energy effective theory will be ordinary flat space.

  What we should exclude is the possibility that all space points form
an aggregate which occupies only a small fraction of the available volume.
But this possibility can be ruled out by a standard entropy argument.
(The same argument explains why the density of gas molecules  is constant
throughout the container which holds them).
Suppose that ground state were an aggregate whose size
is smaller than $L/2$. The relative phase space for such configurations is
$2^3/2^N$, where the numerator counts the eight different regions in which the
aggregate may be localized. Moreover, since the matter action is diffeomorphism
invariant, the change in the matter free energy that arises from confining
all space points to a small region will be negligible.
The relevant triangulations will be distinguishable from ones which occupy
evenly the entire volume only by surface effects which, in turn, will be
suppressed in the limit of large $N$. The leading $N$-dependence of the
effective measure for all such configurations therefore goes
like $2^{-N}$, and so they are completely suppressed in the continuum limit.

  A more exotic possibility is that some structure may form which occupies
the entire volume, but whose unit cell has, say, a fivefold symmetry. In this
case, although the density of points will be approximately constant,
the ground state would not be invariant under a discrete translation group.
We are unable to make further comments on this situation at the moment, except
to express hope that this will not turn out to be the generic situation.

  Finally, we note that even if none of the spacetime symmetries of the
DSRT lattice are spontaneously broken, this does not yet guarantee recovery
of full Lorentz covariance. This is the price that we have to pay for the
special role of time. We return to this issue in Sect.~3.

\vspace{2ex}
\noindent {\bf 2.4~~The zero momentum excitation}
\vspace{1ex}

  Being a lattice model, the continuum limit of the partition function
eq.~(2.2) should correspond to  $a_0 M\to 0$ and
$\ba M\to 0$, with the product $N_0 a_0$ held fixed.
The precise behaviour of the ratio $\ba/a_0$ in the continuum limit
will be discussed in Sect.~3. We will now prove the following statement.
{\it In the limit $a_0 M\to 0$ and regardless of the value of $\ba M$,
the model defined by the partition function}~(2.2) {\it and the action}~(2.5)
{\it has a stable zero momentum one particle excitation, whose
energy is given by $E=M$ exactly}.

  The proof is based on the bilinearity of the lagrangian~(2.6) and the fact
that, for $M=0$, the scalar action is invariant under a global shift symmetry
$$
  \F_{n,t} \to \F_{n,t} + \a \,,
\eqno(2.20)
$$
where $\a$ is a constant. Notice that the measure is invariant under the
{\it local} shift
$$
  \F_{n,t} \to \F_{n,t} + \a_{n,t} \,.
\eqno(2.21)
$$

  We proceed by writing down the Ward identity that corresponds to
the broken shift symmetry. Starting from the expectation value of
$\F_t(\e,x^i)$ eq.~(2.10), we perform the local change of variables~(2.21).
The resulting Ward identity is
$$
  \Big< \!\!\Big< \d S\,\, \F_t(\e,x^i) \Big> \!\!\Big> =
  \Big< \!\!\Big< \d \F_t(\e,x^i) \Big> \!\!\Big> \,,
\eqno(2.22)
$$
where
$$
  \d\F_t(\e,x^i) = {3\ba^3\over 4\p\e^3} \sum_{n\in\cd(x;\e)} \a_{n,t} \,.
\eqno(2.23)
$$
The local variation of the action is
$$
  \d S= a_0 \sum_{t=-N_0}^{N_0-1} \d\cl_t \,,
\eqno(2.24)
$$
where
\beqn{2.25}
  \d\cl & = & \ba^3 \sum_n \dtd\a_n \dtd\F_n \NON
        & & + \ba \sum_{\langle mn \rangle} (\a_m-\a_n) (\F_m-\F_n)\NON
        & & + M^2 \ba^3 \sum_n \a_n \F_n \,.
\eeqn
The discrete time derivative $\dtd\a_n$ is defined analogous to eq.~(2.8).

  The Ward identity~(2.22) is not directly useful, because in order to
evaluate it one has to specify the values of the local function $\a_{n,t}$ in
some reasonable way, and to carry out the integrations over all dynamical
variables. But there is one case where the computation simplifies
considerably, and all integrations can be carried out exactly.

  We now assume that $\a_{n,t}=\a_t$ is a function of $t$ only. Moreover,
we replace the local operator $\F_t(\e,x^i)$ by $\tilde\F_0$, where
$\tilde\F_t$ is the zero momentum projection
$$
  \tilde\F_t = N^{-1} \sum_n \F_{n,t} \,,
\eqno(2.26)
$$
evaluated at time $t$. For this particular choice, the second term on the
\rhs of eq.~(2.25) vanishes, and the Ward identity simplifies to
$$
  \sum_{t=-N_0}^{N_0-1} \left\{
     \dtd\a_t ( G_{t+1} - G_t ) + a_0 M^2 \a_t G_t \right\} =  \a_0 \,.
\eqno(2.27)
$$
Here $G_t$ is the correlator
$$
  G_t = L^3\, \Big< \!\!\Big< \tilde\F_t\,\, \tilde\F_0 \Big> \!\!\Big> \,.
\eqno(2.28)
$$

  We now let $\a_t = \exp (-i\o a_0 t)$, where $\o$ is one of the allowed
frequencies of the time lattice. We find
$$
  \left( {4\over a_0^2} \sin^2 (a_0 \o/2) + M^2 \right) G(\o) = 1 \,,
\eqno(2.29)
$$
where
$$
  G(\o) = a_0 \sum_{t=-N_0}^{N_0-1} e^{-i\o a_0 t} G_t \,.
\eqno(2.30)
$$

  Eq.~(2.29) is an exact expression for the zero momentum correlator.
Taking the limit $a_0\to 0$ and analytically continuing to Minkowski space we
find that the zero momentum propagator has a pole at $E=M$. We have thus
arrived at the remarkable result, that the parameter $M$ in the
lagrangian~(2.6) is the exact energy of the zero momentum scalar excitation.
We also notice that this is a stable excitation, because the action is
invariant under the discrete transformation $\F \to -\F$, and
so the number of $\F$-particles is conserved modulo two.

  This result provides the first indication that the DSRT model possesses a
non-trivial continuum limit. More detailed analysis will be needed in order
to prove that the zero momentum excitation is the end point of a continuous
spectrum, and not just an isolated state. This might be achieved by the
application of the above Ward identity to localized
states, together with the use of some variational estimates.

  We would like to stress, however, that even the present result contains
some non-trivial information about the dynamics of the model.
The relation $E=M$ implies that the zero momentum component of the scalar
field decouples from spacetime dynamics. Showing that the zero
momentum state is the end point of a continuous spectrum amounts to
showing that this decoupling occurs gradually as the scalar state becomes
more and more delocalized. But apriori it is not at all obvious that even the
zero momentum state should decouple from the dynamics of the random lattice.

  To illustrate how things could be different let us consider, instead of the
diffeomorphism invariant lagrangian~(2.6), an action of the kind studied in
\rf{col}. There, the aim was to provide a non-perturbative definition of
quantum field theories in flat space by using a dynamical random lattice.  The
guiding principle in the construction of the  action, was that its classical
continuum limit should coincide with the corresponding {\it flat space}
continuum action. Such an action must depend {\it explicitly} on the embedding
of the triangulations in the target space.  For example, one has to account for
the length of each link and for the volume of cells in the dual lattice. Thus,
while the action is still bilinear in the scalar field, it has terms that
depend in a complicated non-polynomial way on the coordinates of the spacetime
points. Since that action is not bilinear in the dynamical degrees of
freedom, one cannot prove the existence of a non-trivial continuum limit using
the present approach. In fact, it is not clear at all that a non-trivial
continuum limit exists in that model.

\vspace{2ex}
\noindent {\bf 2.5~~The spacetime-induced interaction}
\vspace{1ex}

  Before we turn to discuss how gravity may show up in the continuum limit
of the DSRT model, there is one more interesting result which
can be proved rigorously. To this end, we compare the partition function~(2.2)
of the one scalar  model, with the partition function of a two scalars
model
$$
  Z(2)= \int_{n,t} \cd y \int_{n,t} \cd \F \int_{n,t} \cd \F' \,
        e^{-S(\F)-S(\F')} \,.
\eqno(2.31)
$$
The action $S$ is still given by eqs.~(2.5) and~(2.6). The partition
function~(2.31) describes two scalar fields which are uncoupled accept
possibly through the spacetime dynamics. We now claim that, in fact, spacetime
fluctuation induce an {\it attractive} interaction among the scalar
excitations.

  The partition function~(2.2) defines an average of the matter free energy
$$
  e^{-F(y_{n,t})} = \int_{n,t} \cd \F\,  e^{-S(\F)} \,,
\eqno(2.32)
$$
with respect to the normalized spacetime measure $\int_{n,t} \cd y$. Let us
denote this average by
$$
  Z = \left< e^{-F} \right> \,.
\eqno(2.33)
$$
It is easy to see that the two fields partition function is
$$
   Z(2) = \left< e^{-2F} \right> \,.
\eqno(2.34)
$$
Since the free energy has {\it some} dependence on the positions of the
spacetime points, the averaging is non-trivial and we obtain
$$
  \left< e^{-2F} \right>   >    \left< e^{-F} \right>^2 \,.
\eqno(2.35)
$$
Let us introduce a dimensionless vacuum energy density through
$$
  \ce = -1/(N N_0) \log Z \,,
\eqno(2.36)
$$
Analogous definition applies to $\ce(2)$, the vacuum energy density of the two
scalars model. Eq.~(2.35) implies
$$
  \ce(2) < 2\ce \,.
\eqno(2.37)
$$

  Eq.~(2.37) is the main result of this subsection. If spacetime fluctuations
did not induce any interaction among the matter fields, we would expect that
$\ce(2)$ should be equal to $2\ce$, as in the case of two uncoupled scalar
fields on a regular lattice. From  the fact that $\ce(2)$ is actually lower
than $2\ce$, we learn that spacetime fluctuations induce an {\it attractive}
interaction among matter excitation.

  This result is in fact very general. The one condition
needed to establish inequality~(2.35) is that the action should depend only on
translationally invariant  properties of the spacetime triangulations.
Thus, it applies to a much more general class of actions and not only to
diffeomorphism invariant ones. While it is encouraging to find that the
generic spacetime-induced interaction is attractive, one expects that
the details of that attractive interaction should be model dependent.
One needs a more quantitative information about that interaction in order to
tell for which models (if any) it gives rise to a long range force in the
continuum limit.


\vspace{5ex}
\noindent {\bf 3.~~INDUCED GRAVITY}
\vspace{3ex}

  In this section we investigate the feasibility
of obtaining the familiar gravitational interaction in the continuum limit
of the DSRT model. We begin in subsec.~3.1
by calculating the classical continuum limit of the scalar action. To
facilitate this calculation, we assume that the following relation
$$
  \F_{n,t}^\l = \left.\F^{\l'}(x,t)\right|_{x=y_{n,t}}\,,\quad\quad \l\ll 1 \,,
\eqno(3.1)
$$
holds between low energy eigenstates $\F_{n,t}^\l$ of the discrete action,
and smooth continuum wave functions $\F^{\l'}(x,t)$. Here $\l$ and $\l'$
denote resprectively the eigenvalues of the discrete and continuum actions,
and what will be described below can be thought
of as a self-consistent procedure to determine $\F^{\l'}(x,t)$.

  In order to describe the dynamics of the model, we introduced in sect.~2
a set of observables that depend on the embedding of the
discrete spacetime structure in the target flat space.
If this is an adequate set of observables, then it should
be possible to extract the dynamical curved metric felt by low energy
observers by enforcing the correspondence~(3.1) on the discrete matter action.

  Taking the classical continuum limit of the matter action necessitates
the introduction of two thermodynamical fields, a symmetric tensor
and a field that describes the local density of points. We recover
the continuum action of a scalar field in a gravitational background, where the
symmetric tensor plays the role of the inverse curved metric, and provided
one can identify the density field with the spatial volume element of
the curved metric. Whether or not the fluctuations of the local point density
coincide with the fluctuations of the volume element of the curved metric
tensor, is a dynamical question. Later we will argue that this should be the
case in the continuum limit.

   Because of the preferred role of time, only inverse metrics with $g^{00}=1$
are attainable in the DSRT model. From the point of view of the curved space
effective theory at low energies, this can be interpreted as partial
gauge fixing. Another property which can be interpreted that way, is the local
conservation law of space points (see subsect.~3.4).

  The partition function of the DSRT model defines an effective measure for
the gravitational (and density) fields.  In subsect.~3.2 we introduce an
improved spacetime measure which involves an integration over a small subset
of the diffeomorphism group.  We then argue that, in the continuum limit, the
effective gravitational measure reduces to the product of a
gauge fixing condition which is natural for the model, times the exponential
of an effective action which is invariant under a
subgroup of infinitesimal general coordinate transformations (eq.~(3.17)
below). This invariance is sufficient to ensure the masslessness of the
graviton.

  The rest of this section is devoted to a discussion of the expected
properties of the effective gravitational action. While the discussion
is rather heuristic at this stage, it indicates that satisfactory solutions
to key issues in Quantum Gravity are feasible in the DSRT model.
Of course, only detailed investigations will
ultimately tell whether or not the physical picture that we present here
is correct. But one cannot carry out any detailed study of the model without
first having in mind some idea of what one is looking for.

  In subsect.~3.3 we show that the effective action cannot contain a
cosmological term. Next, we argue that the Planck scale should be
identified with the largest mass of the matter sector.
We propose the name {\it pregraviton} to denote any ordinary particle
of Planckian mass. In our model, gravitons
are excitations of the quantum spacetime, whose {\it size}
is determined by the Compton wave length of the  pregraviton.
We discuss the possibility that perturbative processes above the Planck scale,
but still much below the cutoff scale, may be described by an
{\it asymptotically free}, curvature squared continuum theory~\cite{ftz}.
Finally, in subsect.~3.4 we discuss in what ways, and under
what circumstances, the physics of the DSRT model may differ from General
Relativity.

\newpage
\vspace{2ex}
\noindent {\bf 3.1~~The classical continuum limit of the matter lagrangian}
\vspace{1ex}

  Because time is discretized into regular intervals, taking the continuous
time limit of the summation on the \rhs of eq.~(2.5) is trivial.
What we have to do, is to calculate the classical continuum limit of the
lagrangian~(2.6) by enforcing the correspondence~(3.1) on the scalar
field's configurations. On the \rhs of eq.~(3.1), the coordinates $x$
and $t$ take values in the flat four-torus $L^3\times [0,2a_0 N_0]$. As we
will see, an effective curved metric is induced dynamically in that flat space.

  In deriving the classical continuum limit we introduce a coarse graining
scale $l$. We will assume that  $\F^{\l'}(x,t)$ is
slowly varying inside a  ball of radius $l$.
To justify a thermodynamical treatment of the spacetime triangulations, we
demand that $l\gg \max\{a_0,\ba\}$.  Since $l$ is arbitrary, we want that
a change in $l$ will amount to finite renormalization of the parameters of
the effective low energy theory. To make this possible, we assume that $l$
is much smaller then any physical distance scale in the theory.

  In the continuum limit both $a_0 M$ and $\ba M$ should tend to zero.
As we will see, requiring consistency of the analytic continuation to
Minkowski space, implies that in the continuum limit the ratio $\ba/a_0$
should tend to zero too.

  We begin with the mass term which is the simplest one to consider.
We make use of the identity
$$
  1 = {3\over 4\p l^3} \int d^3x\, \th(x-y;l) \,,
\eqno(3.2)
$$
where
$$
  \th(z;l) = \left\{\begin{array}{ll}
     1\,,\quad &  |z| \le l \,, \\
     0\,,\quad &  |z|  >  l \,.
                              \end{array}\right.
\eqno(3.3)
$$
Substituting the identity~(3.2) in the mass term of the lagrangian~(2.6) and
interchanging the order of summation and integration, we find
(suppressing the time dependence of the dynamical variables)
\beqn{3.4}
 \cl_m & = &
       \half M^2 {3\ba^3\over 4\p l^3}\sum_n\int d^3x\,\th(x-y_n;l) \F_n^2 \NON
       & = &
       \half M^2 {3\ba^3\over 4\p l^3} \int d^3x \sum_{n\in\cd(x,t)} \F_n^2 \,,
\eeqn
where $\cd(x,t)\equiv\cd(x,t;l)$ is the ball in the time slice $t$ with center
at $x$ and radius $l$. In eq.~(3.4) we do not worry about points that happen
to sit on the boundary of $\cd(x,t)$. Also, when evaluating the hopping terms
below, we will not worry about links that cross that boundary. The prescription
chosen for handling these cases in unimportant, as long as it respects
the translation and rotation invariance of the model.
A simple prescription would be for example to include in the sum only points or
links which lie entirely inside a given ball.

  Approximating $\F_n$ by its value at the center of $\cd(x,t)$ we obtain
$$
  \cl_m= \half M^2 \int d^3x\, \r(x,t) \F^2(x,t)\,,
\eqno(3.5)
$$
where
$$
  \r(x,t) = {3\ba^3\over 4\p l^3}\, n(x,t) \,.
\eqno(3.6)
$$
In the above equations, $n(x,t)$ is the number of lattice points inside the
ball $\cd(x,t)$, and $\r(x,t)$ is a rescaled, dimensionless density of points.

  We next turn to the gradient term (the second term on the \rhs of eq.~(2.6)).
For nearest neighbour sites inside the ball $\cd(x,t)$ on a given time slice,
we make the approximation
$$
  \F_m - \F_n = (y_m^i - y_n^i)\, \partial_i \F(x,t)\,.
\eqno(3.7)
$$
Proceeding along the same lines as before we find
$$
  \cl_{grad} = {3\ba\over 8\p l^3} \int d^3x  \sum_{ {\langle mn \rangle}
               \in\cd(x,t)}
      (y_m^i - y_n^i)(y_m^j - y_n^j) \partial_i \F(x,t) \partial_j \F(x,t)\,.
\eqno(3.8)
$$
Introducing the dimensionless tensor field
$$
  \trg^{ij}(x,t)= {1\over \ba^2 n(x,t)} \sum_{ {\langle mn \rangle}
                  \in\cd(x,t)}
             (y_m^i - y_n^i)(y_m^j - y_n^j) \,,
\eqno(3.9)
$$
we finally obtain
$$
  \cl_{grad} = \half \int d^3x\, \r \trg^{ij}\, \partial_i \F \partial_j \F\,.
\eqno(3.10)
$$
As suggested by eq.~(3.10), $\trg^{ij}(x,t)$ will be identified with the
inverse of the space-space part of the metric tensor.

  We last turn to the kinetic term. The discrete time derivative~(2.8) involves
the world line $Y_{n,t}$. In calculating its continuum analog, we have to take
into account both the explicit time dependence of $\F_{n,t}$ and its implicit
time dependence through the motion of $Y_{n,t}$. We thus arrive at
$$
  \dtd\F_{n,t} = \left.\left( \dt\F(x,t) +\dt{Y}^i_{n,t}\,
    \partial_i \F(x,t) \right) \right|_{x=Y_{n,t}} \,.
\eqno(3.11)
$$

  In the above equation we have substituted $\dt{Y}_{n,t}$ for
$\dtd{Y}_{n,t}$. This is not really an additional approximation.  To justify
eq.~(3.11) already for $\dtd{Y}_{n,t}$
the displacement $a_0 \dtd{Y}_{n,t}$ should be small compared to the scale
of variation of $\F(x,t)$ and, in any event, it should not be allowed to
become arbitrarily large.

 There are two effects that govern the world line $Y_{n,t}$. One is the
statistical fluctuation due to the random nature of the spacetime measure.
The other is a collective motion that may be induced, say, by a macroscopic
matter distribution. Thanks to our time linking prescription,
the fluctuating part of $\dtd{Y}_{n,t}$ is $O(\ba/a_0)$.
In order to justified eq.~(3.11) we demand that the continuum
limit should correspond to $\ba/a_0\to 0$. In this limit, only the smooth,
collective component of each world line survives, and so in this limit the
world line is {\it differentiable}.

   By contrast, had we taken the opposite limit, we would have found that
after $\ba/a_0$ time slices (which would be a very big number), the world line
has been carried away by brownian motion to a macroscopically large distance
compared to $\ba$. Hence, this limit is unacceptable. For example, it is likely
that in this limit, localized states may dissipate into completely
delocalized ones in a finite time.

  Substituting eq.~(3.11) in the kinetic term we now find
$$
  \cl_k = \half \int d^3x\,\r\, \tmg^{\m\n}\,\partial_\m \F \partial_\n \F\,.
\eqno(3.12)
$$
The components of the symmetric tensor $\tmg^{\m\n}(x,t)$ are
\beqrabc{3.13}
 \tmg^{00}(x,t) & = & 1\,, \\
 \tmg^{0i}(x,t) & = & n^{-1}(x,t) \sum_{n\in\cd(x,t)} \dt{Y}^i_n \,, \\
 \tmg^{ij}(x,t) & = & n^{-1}(x,t) \sum_{n\in\cd(x,t)} \dt{Y}^i_n \dt{Y}^j_n \,.
\eeqr
Notice that $\tmg^{0i}(x,t)$ describes a {\it collective} drift velocity of
the spacetime medium. Because of the coherent character of that motion,
one has $\tmg^{ij} = \tmg^{0i} \tmg^{0j}$. Henceforth we will assume this
relation to hold except under extreme circumstances (see subsect.~3.4).

  The last step is to introduce the symmetric tensor
$\frg^{\m\n}(x,t)$ which is identified with the inverse of the curved
four-metric. Its components are
\beqrabc{3.14}
   \frg^{00}(x,t) & = & 1\,, \\
   \frg^{0i}(x,t) & = & \tmg^{0i}(x,t) \,, \\
   \frg^{ij}(x,t) & = & \trg^{ij}(x,t)+\tmg^{ij}(x,t) \,.
\eeqr
Eq.~(3.14) resembles the ADM parametrization of the inverse metric~\cite{mtw},
with the lapse function set equal to one. But, in the DSRT model eq.~(3.14) is
more than a parametrization: it defines the gravitational field in terms of
the microscopic, quantum  spacetime structure. The metric tensor
$\frg_{\m\n}(x,t)$ itself is
\beqrabc{3.15}
   \frg_{00}(x,t) & = & 1 + \tmg^{0i} \tmg_{0i}\,, \\
   \frg_{0i}(x,t) & = & -\tmg_{0i}(x,t) \,, \\
   \frg_{ij}(x,t) & = & \trg_{ij}(x,t) \,.
\eeqr
On the \rhs of eq.~(3.15) $\trg_{ij}$ is the inverse of $\trg^{ij}$, and
space indices are lowered and raised using the three-metric. (The position of
the index $0$ in $\tmg$ is unimportant).

  A formal analytic continuation of the curved metric to Minkowski space is
facilitated by substituting $-1$ instead of $1$ on the \rhs of eq.~(3.15a).
As we have explained earlier,  our continuum limit forces the time-space
components $\tmg^{0i}$ to describe a collective motion of the space
points. Under most circumstances, we expect this motion to vanish or to be
negligibly small. We now see that both the collective character and the
smallness of that motion are indeed necessary, in order that the effective
curved metric will not become singular and will have the correct signature
after the analytic continuation.

  Expressed in terms of the gravitation and density fields, the continuum
action is
\beqn{3.16}
  S_{cont} & = & \int dt\, \cl_{cont} \NON
           & = & \half\int dt d^3x\, \r\,
      \left\{ g^{\m\n} \partial_\m \F \partial_\n \F + M^2 \F^2 \right \} \,.
\eeqn
This is recognized as the continuum action for a massive scalar field in the
background of an external gravitational field, provided we can identify the
density field $\r(x,t)$ with the curved volume element. Notice that, since
$g^{00}=1$, the local four-volume element is equal to the local three-volume
element~\cite{mtw}.

  From the point of view of the low energy effective curved space, $g^{00}=1$
looks like a partial gauge fixing of general coordinate invariance.
There is a large subgroup of  coordinate transformations which
leave the component $g^{00}$ invariant. These are given by
\beqn{3.17}
  t' & = & t \,, \NON
  x' & = & x'(x,t) \,.
\eeqn
The coordinate transformation~(3.17) will play an important role below.

  Whether or not one can identify the density field $\r(x,t)$ with
$\sqrt{g(x,t)}$ is a dynamical question. It is convenient to examine this
question by first making the {\it field redefinition}
$$
  \r(x,t) = e^{2\f(x,t)}\, \sqrt{g(x,t)}\,.
\eqno(3.18)
$$
If eq.~(3.18) is substituted in eq.~(3.16), we obtain the generally covariant
continuum action of a scalar field in a dilaton-gravity background, where
$\f(x,t)$ is the ``dilaton'' field. This provides a convenient parametrization
to study the dynamical questions. Namely, does the gravitation field as
defined above propagates massless excitations, and does the ``dilaton'' field
propagates any low energy excitations. These questions are addressed in
the following subsection.

  We will argue that, with a certain improvement of the spacetime measure,
the curve metric that we have introduced should have massless spin two
excitations in the continuum limit. As for the ``dilaton'' field, we find
that this field is completely frozen. As a result, the dynamics of the DSRT
model justifies the identification of $\r(x,t)$ with $\sqrt{g(x,t)}$.

  We close with a comment on the expectation values of the thermodynamical
fields. In subsect.~2.3 we argued that the ground state as probed by
low energy observers, should be Poincar\'e invariant. Neglecting possible
microscopic crystal-like structure, which is irrelevant for most
of the subsequent discussion, we thus have constant expectation values for
the thermodynamical fields. The expectation value of the density field is
trivially found to be $\svev{\r(x,t)} = 1$. The expectation value of the
metric tensor is diagonal, and by making some redefinition of parameters
we may assume that it takes the canonical value
$\svev{g^{\m\n}(x,t)}= \d^{\m\n}$.

\vspace{2ex}
\noindent {\bf 3.2~~An improved spacetime measure}
\vspace{1ex}

  In subsect.~3.1 we identified the components of the curved metric tensor
with quasi-local averages of certain microscopic properties of the
discrete spacetime structure. If this is a valid identification, then the
thermodynamical curved metric field should be governed by
Einstein's equation in the appropriate limit.

  In order to investigate our dynamical curved metric we define
an {\it effective measure} $\O(g^{\m\n},\f)$ as follows
$$
  \O(g^{\m\n},\f) =  \left\langle\!\!\left\langle
                       e^{-F} \right\rangle\!\!\right\rangle_{g^{\m\n},\f} \,.
\eqno(3.19)
$$
The expectation value on the \rhs of eq.~(3.19) is a partial ensemble average
in the spacetime measure $\int_{n,t} \cd y$,
which is taken over all triangulations that
correspond to given $g^{\m\n}$ and $\f$ fields. $F$ is the matter free energy.
If we are interested in an energy range below some physical scale, then we have
to include in $F$ only the contribution of matter excitations with energies
comparable to or larger than that scale. The infra-red cutoff of the matter's
functional integration can be imposed self-consistently, because low energy
matter eigenstates should depend only on the thermodynamical fields (and
not on the microscopic details) and their energies are invariant with respect
to the coordinate system chosen to describe the effective low energy
curved space.

  The space of metrics $g^{\m\n}(x,t)$ over the four-torus has a natural
fiber bundle structure, where each fiber is an equivalence class of all
metrics which are obtained from some initial metric by a general coordinate
transformation. The gravitation field will have the desired dynamics if,
in the continuum limit, the effective measure takes the form
$$
  \O(g^{\m\n},\f) = \d(\cc(g^{\m\n},\f))\, e^{ - S_{eff}(g^{\m\n},\f)}\,.
\eqno(3.20)
$$
In eq.~(3.20) $S_{eff}(g^{\m\n},\f)$ is a generally covariant effective
action, and $\d(\cc(g^{\m\n},\f))$ stands for a generic  (possibly
partial) gauge fixing condition.

  In other words, there is no need that the support of the effective
measure will cover all representatives of each metric.
The crucial requirement is that, inside each fiber, the effective measure
should be constant over its support. The saddle point of the effective
measure will then be a solution of a generally covariant equation of motion.

  Our definition of the curved metric depends on the flat coordinates on the
four-torus. Because of the statistical character of our $g^{\m\n}(x,t)$,
we  expect that in each fiber, the effective gravitational measure
will be highly peaked around those representatives
where $g^{\m\n}(x,t)$ is as close as possible to a constant function.
The reason is simply that the phase space for such configurations is maximal.
Moreover,  the dynamical geodesic distance
inferred from these representatives should reduce to the
coordinate distance in the underlying flat space, provided one is
sufficiently far away from matter concentrations. This is a good feature
that we should not spoil unnecessarily.

  In this subsection we  introduce an improved spacetime measure, which
incorporates an integration over a small subset of the diffeomorphism
group.  We will argue that, with the improved measure,
eq.~(3.20) should hold in the continuum
limit, where $S_{eff}(g^{\m\n},\f)$ is invariant under {\it infinitesimal}
general coordinate transformations of the form~(3.17). While we will not
bother to calculate the details of the effective gauge fixing condition, we
know that it selects representatives with the qualitative properties described
above.

  In order to define the improved spacetime measure we introduce an auxiliary
cubic lattice which is overlayed with the flat target space from which the
points on every time slice are taken. The infrared scale $L$ is assumed to be
an integer multiple of $a'$, the lattice spacing of the cubic lattice.
Like $a_0$ and $\ba$, this is a cutoff parameter which has to be sent to zero
in the continuum limit. Its physical role (see below) suggests that we should
take $a'/\ba\gg 1$. We already have one cutoff parameter with this property,
namely $a_0$. The natural choice is to take $a'=a_0$. Henceforth we will
denote the common value of these two parameters by $a$.

  The improved spacetime measure contains a new degree of freedom,
a {\it spatial} vector field $\x_{x,t}^i$ that resides on the sites of the
cubic lattice. We will use the convention that when the coordinate $x$ appears
as a subscript, it takes values on the sites of the auxiliary regular
lattice. Finally, there is also an additional dimensionless
parameter $\d$ which serves to determine the subset of the diffeomorphism
group over which we are integrating. As we will see, $\d$ can be sent to zero
in the continuum limit, but rather slowly compared to other cutoff parameters.

  The new partition function is defined by
\beqn{3.21}
  Z_D  & = & Z(N_0,N,L,a,\d,M) \NON
       & = & \int_{n,t} \cd y  \int_{n,t} \cd \F
             \int_{x,t} \cd \x\, e^{-S}\,,
\eeqn
where
$$
  \int_{x,t} \cd \x \equiv \cn (a\d)^{-6N_0 (L/a)^3}\, (a\d/L)^3
         \prod_{x,t} \int d^3 \x_{x,t}\, \cf_\d(\x).
$$
Notice that $2N_0 (L/a)^3$ is the number of sites of the auxiliary lattice.
The name $Z_D$ is chosen to remind us that the improved measure has certain
built in diffeomorphism invariance. The function $\cf_\d(\x)$ determines the
range of the vector field $\x_{x,t}$ and, hence, the subset of the
diffeomorphism group over which will are integrating. We take
\beqn{3.22}
  \cf_\d(\x) & = & \prod_{x,t,k}
               \th \Big( | (x,t)' - (x+\hat{k},t)' | - a \,;\, a\d \Big) \NON
    & &  \times \prod_{x,t}
               \th \Big( | (x,t)' - (x,t+1)' | - a \,;\, a\d \Big) \,,
\eeqn
where $\th$ has been defined in eq.~(3.3), and
$$
  (x^i,t)' = (x^i + \x_{x,t}^i,t) \,.
\eqno(3.23)
$$
Notice that eq.~(3.23) is a discretization of the coordinate
transformation~(3.17),
thought of as an actual deformation of spacetime. $x+\hat{k}$ is the
neighbouring site of $x$ in the positive $k$ direction. The meaning of these
definitions is that under the (as yet discrete) transformation~(3.23), the
fractional change in the distance between the images of two neighbouring
sites should not exceed $\d$.

  In spite of the introduction of the regular lattice, the new
partition function $Z_D$ has the same global spacetime symmetry as the old one.
Continuous translation invariance is restored by the integration over the
diffeomorphism's translational zero mode. Likewise, in the new partition
function, rotation invariance is broken only by the infrared boundary
conditions. The normalization of the  $\int_{x,t} \cd \x$ measure takes into
account that the basic range of variation of $\x^i_{x,t}$ is $O(a\d)$, and the
factor $(a\d/L)^3$ compensates for the unrestricted range of the zero mode of
$\x^i_{x,t}$. The normalization constant $\cn$ is therefore $O(1)$. We will
not need the numerical value of $\cn$ and, for most purposes we can simply
drop it from the definition of $Z_D$. Notice that, to derive inequality~(2.37)
for the improved measure we only have to know that $\cn$ exists.

  The triangulations are built in two steps. We first connect a given set of
spacetime points $\{y^i_{n,t}\}$ according to the prescription of Sect.~2.
The {\it embedding} of the triangulation in the target flat space is
then determined by the vector field $\x^i_{x,t}$.

  To proceed, each vector field configuration $\x^i_{x,t}$  is first extended
to a smooth, continuous vector field $\x^i(x,t)$. We require that the vector
field $\x^i(x,t)$ should coincide with the original vector field configuration
$\x^i_{x,t}$ when $(x,t)$ is the coordinate of a site on the auxiliary lattice.
The diffeomorphism generated by $\x^i(x,t)$ is then applied to the entire
triangulation. Namely, without changing the connectivity of the triangulation,
we move the lattice points to new locations given by $$ y^i_{n,t} \to
(y')^i_{n,t} = y^i_{n,t} + \x^i( y_{n,t},t ) \,.  \eqno(3.24) $$

  What is described above is a family of embeddings, or deformations,
of {\it the same} triangulation. These deformations leave the matter action
invariant, and so the improved partition function really defines the same
quantum theory as the original one. In fact, with the normalization constant
$\cn$ included, one has $Z_D=Z$. The advantage of the improved measure
it that the relevant observables (the ones defined in analogy to subsect.~2.2)
are local. The local observables of the improved partition function are
related to a set of non-local observables of the original partition function.

  We will now show that the embeddings described by eq.~(3.24)
{\it generate a curved metric tensor and a density field which are related
to the original ones at $\x^i_{x,t}=0$ by a general coordinate
transformation of the special type}~(3.17).

  To find how the new metric looks we have to apply the transformation~(3.24)
in eqs.~(3.9) and~(3.13). Let us start with the spatial part $\trg^{ij}$
(eq.~(3.9)). Suppressing the $t$-dependence, we make the following
approximation for any two points which are connected by a space link
\beqn{3.25}
  (y')^i_n - (y')^i_m & = & (y^i_n - y^i_m) +
        ( \x^i(y_n) - \x^i(y_m) )  \NON
     & = & (y^j_n - y^j_m) (\d^i_j  + \partial_j \x^i) \,.
\eeqn
Substituting in eq.~(3.9)  we find
\beqn{3.26}
  (\trg')^{ij}(x',t) & = &  (\d^i_k  + \partial_k \x^i)
                           (\d^j_l  + \partial_l \x^j) \times \NON
     & &  {1\over \ba^2 n'(x,t)} \sum_{ {\langle mn \rangle}   \in\cd'(x,t)}
              (y_m^k - y_n^k) (y_m^l - y_n^l)\,.
\eeqn

  The \rhs of eq.~(2.26) is evaluated on the original triangulation. But
this is not yet an expression for the transformed metric in terms of the
old one. The reason is that the summation is not carried over the links
that belong to $\cd(x,t)$. Rather, it is carried over the links that belong to
$\cd'(x,t)$, the {\it pullback} of $\cd(x',t)$. Similarly, $n'(x,t)$ is
the number of  points in $\cd'(x,t)$. The pullback $\cd'(x,t)$
is defined as
$$
  \cd'(x,t) = \Big\{ (z,t)\, \Big| \,(z',t) \in \cd(x',t) \Big\} \,.
\eqno(3.27)
$$

  The typical deformation of the pullback $\cd'(x,t)$ compared to $\cd(x,t)$
is of order $a\d$. In the continuum limit we let
$\ba/a\to 0$ and so we expect that, already on the scale $a$, the
thermodynamical fields will fluctuate very little around their mean values.
Moreover, as long as we perturb the ground state only by smooth, extended
sources, we anticipate these mean values to be slowly varying. We can therefore
self-consistently replace the primed quantities on the \rhs
of eq.~(3.26) by unprimed ones. We thus obtain
$$
  (\trg')^{ij}(x',t) = (\d^i_k  + \partial_k \x^i)
                    (\d^j_l  + \partial_l \x^j) \trg^{kl}(x,t) \,.
\eqno(3.28)
$$
Making a similar analysis for the transformation of $(\tmg')^{\m\n}$ we get
$$
   (g')^{\m\n}(x',t) = ( \d_\s^\m + \partial_\s \x^\m)
                   ( \d_\t^\n + \partial_\t \x^\n) g^{\s\t}(x,t) \,.
\eqno(3.29)
$$
In order to write this compact transformation law we have introduced a
vanishing zeroth component field $\x^0(x,t)=0$. Finally, one can check
that the density field $\r(x,t)$ transforms as a scalar density, namely
$$
  \r'(x',t) = \det^{-1}(\d^i_k  + \partial_k \x^i)\, \r(x,t) \,.
\eqno(3.30)
$$
This completes the demonstration that the effect of the $\x$-transformations
on the gravitation and density field is equivalent to the effect of a
general coordinate transformation.

  To see how the integration over the $\x$-field enforces the form~(3.20)
in the continuum limit, let us consider the width of $\O(g^{\m\n},\f)$
inside each fiber, as derived from the original partition function eq.~(2.2).
The gravitation field is obtained by a statistical average over a region of
size $l$. The lower bound on the size of that region is $a$, and therefore
the number of points involved in this average is at least
$n_0\approx (a/\ba)^3$. In the continuum limit $n_0$
tends to infinity. We thus expect $\O(g^{\m\n},\f)$ to be
highly peaked around a mean value, with a statistical uncertainty which is
governed by $n_0^{-\g}$ for some calculable $\g>0$. In the continuum limit,
$\O(g^{\m\n},\f)$  will rapidly tend to a delta function.

  Turning to the effective measure $\O_D(g^{\m\n},\f)$ of the improved
partition function, we now see that, inside each fiber,
is it obtained by a convolution of $\O(g^{\m\n},\f)$ over some domain
whose size is $O(\d)$. If we approach the continuum limit in a controlled
way, such that $\d$ is kept very big {\it relative} to  $n_0^{-\g}$,
then the improved effective measure will be almost constant over its
support inside each fiber. As we move inside a given fiber, the width of
$\O(g^{\m\n},\f)$ will now govern only the width of the tiny
boundary region over which $\O_D(g^{\m\n},\f)$  abruptly changes from its
constant positive value to zero.

  In the continuum limit, while $\d$ tends slowly to zero, $\d/n_0^{-\g}$
tends to infinity, and so the limiting effective measure inside each
fiber becomes exactly constant over its support.
This gives rise to the desired form~(3.20) for the
effective gravitational measure, where the effective action is invariant
under infinitesimal coordinate transformations~(3.17), and the
effective gauge fixing condition
has the desirable features described in the beginning of this subsection.

  Invariance of the effective action under infinitesimal transformations
of type~(3.17) is sufficient to ensure the
masslessness of the graviton. The argument is a standard one.
We define $g^{\m\n}(x,t)=\d^{\m\n}-h^{\m\n}(x,t)$. (Notice the slight
modification compared to the usual definition, which we do because in the
DSRT model $g^{00}=1$ identically, \ie according to our definition
$h^{00}=0$). Let us now write down
all possible mass terms for $h^{\m\n}$. We have to allow time and
space components to appear in an asymmetric way, because time and space are
not treated symmetrically in the DSRT lattice. In fact, we know that the
spin two part is contained in the space-space components of the curved metric,
but we might as well check for possible mass terms for the time-space
components. The only possible mass terms are $(h^{0k})^2$, $h^2$ and
$(h^{kl} - {1\over 3} \d^{kl} h)^2$, where $h=h^{kk}$. The reader can easily
verify that no linear combination of these terms is invariant under
infinitesimal coordinate transformations~(3.17). Hence, the curved metric
field should have massless spin two excitations.

  As for the ``dilaton'' field, substituting the defining eq.~(3.18) in
the continuum action~(3.16) would suggest the existence of a global shift
symmetry for $\f(x)$ which is {\it not} a part of the coordinate
transformation group. But the spacetime measure (with or without improvement)
does not have an additional symmetry which acts that way on the $\f$ field.
In particular, the diffeomorphisms~(3.24) leave the $\f$ field invariant.
As a result, the fluctuations of the $\f$ field are suppressed in the
continuum limit, and the $\f$ field is frozen to its expectation value.
Returning to eq.~(3.18) we thus see that, while apriori we had to introduce
the density field as an independent variable, the dynamics forces the density
field to coincide with the volume element of the curved metric tensor.

  We conclude with a technical comment. Since $g^{00}=1$, the curved metric
tensor as defined in subsect.~3.1 contains only nine algebraically
independent degrees of freedom. The density field contains one degree of
freedom. Altogether we have ten degrees of freedom, as the metric tensor
should normally have. It is therefore suggestive to try to make a field
redefinition that will replace the constrained
metric tensor and the density field by an unconstrained metric tensor.

  If we focus on the kinetic term in the continuum action~(3.16), then the
desired field redefinition is easily found to be
$$
  \bar{g}^{\m\n} = {\sqrt{g}\over \r} g^{\m\n} \,.
\eqno(3.31)
$$
The above definition implies the equality of $\r g^{\m\n}$ and
$\sqrt{\bar{g}}\, \bar{g}^{\m\n}$. Therefore, this field redefinition brings
the kinetic term to the standard form without having to rely on the dynamics.
However, there is also a mass term, and that term does not take the standard
form in terms of the new metric tensor. Thus, we are unable to trade
algebraically the independent density field with a tenth component of the
metric tensor, because the continuum scalar action
is not conformally invariant.

  This consideration suggests that a similar trick might work for gauge
fields. But the dynamics forces the proportionality factor $\sqrt{g}/\r$ in
eq.~(3.31) to be equal to unity anyway. Hence, the  final result will be the
same. Even if we can work with the modified metric $\bar{g}^{\m\n}$, the
effective action will be invariant only under the restricted coordinate
transformations~(3.17). Therefore, there is no symmetry that prevents the
appearance of a mass term with a divergent coefficient for $\bar{h}^{00}$.
So again we arrive at the same conclusions, namely, that the component
$\bar{g}^{00}$ is completely frozen, and that the dynamical
density of points is given by the curved volume element.

\vspace{2ex}
\noindent {\bf 3.3~~The effective gravitational action}
\vspace{1ex}

  So far, we provided evidence that the ground state of the DSRT model
as probed by low energy observers, is flat space. We also argued that, with
the improved spacetime measure of subsect.~3.2, the set of observables
defined in Sect.~2 should exhibit the presence of a long range
spacetime-induced interaction.  If both of these assertions
are correct, the distance to proving the emergence of the familiar
gravitational interaction is very short.

  The relevance of the DSRT model to Quantum Gravity therefore depends on
one's ability to prove the validity of the
above assertions. We will not attempt to provide the detailed proofs in this
paper. But if the DSRT model provides a consistent model of
Quantum Gravity, then it should be possible to obtain satisfactory solutions
to numerous well known
problems. Our aim in this subsection is to indicate that such answers may
indeed exist. The following discussion is also important in order to
decide what is the best way to proceed with a detailed study of the properties
of the DSRT model.

  The gravitational effective action defined in the previous subsection
may not, in fact, be the best tool to investigate the model. When we speak
about the gravitational action, we have in mind an approximation where one
keeps a finite number of local terms in a derivative expansion of the exact,
necessarily non-local effective action.
Further complications arise because of the need to
disentangle the effective action from the effective gauge fixing
condition (see eq.~(3.20)). But the effective action provides a convenient
and familiar terminology, and so we will continue to use it in this subsection.

  In order that a local effective action will be a useful
tool in the investigation of the model, a minimal requirement is that
the true ground state should appear as a (possibly local)
minimum of that effective action. In subsect.~2.3 we argued that the ground
state should be homogeneous, and this provides a potential explanation for
the absence of a cosmological term in the effective action.
Moreover, one can interpret instability problems such
as the unboundedness from below of the Einstein-Hilbert
action, as an artifact of the truncation leading from the full non-local
effective action to the local approximate one.

  Two central issues which can be  addressed by
computing an approximate gravitational effective action, are the identification
of the Planck scale with some physical scale of the DSRT model, and
the nature of the effective theory at shorter-then-Planck-scale distances.

  The explicit physical mass scales in the partition function~(3.21)
include one mass parameter for each scalar field. Let us assume that a
particular mass parameter, namely $M$, is much bigger then the others. If we
are interested in physics much below the scale $M$ then we should integrate
out the mass-$M$ scalar field.  The scalar action is
diffeomorphism invariant, and it should clearly give rise to a generally
covariant contribution to the effective gravitational action.

  One can get an idea of the outcome by considering the calculation of the
scalar free energy using a continuum regularization technique which respects
general coordinate invariance, such as dimensional regularization. This
computation was done long ago by `tHooft and Veltman~\cite{tv}. Here we are
not interested in the details of their computation but only in the qualitative
features of their result.  The bottom line is that
the {\it local} part of the scalar
free energy may contain all generally covariant local operators with
dimension up to four. Operators of lower dimensions are
multiplied by the appropriate power of $M$.

  The infrared leading term in this expansion is a cosmological term.
As we discussed above, we do not expect any cosmological term to arise
in the complete effective gravitational action of the DSRT model.
In the present terminology, this means that a compensating, negative
cosmological constant should arise from the dynamics of the spacetime
points. This may turn out to be the entropy factor of subsect.~2.3 in
disguise. The anticipated cancellation signals that the presence of matter
should  have an important effect on the infrared limit of the
spacetime dynamics.

  In the absence of a cosmological term, the leading infrared term is the
Einstein-Hilbert action. For this term, we do not see any reason why
the spacetime dynamics should conspire to cancel a contribution coming
from integrating out the heavy scalar field. On dimensional grounds,
the dimension two coefficient of the Einstein-Hilbert action should
be $O(M^2)$. Thus, the Planck scale is identified with the highest mass scale
of the matter sector. We believe that this should be a general relation
which is valid also for other choices of the matter sector. For example, in a
model containing several asymptotically free non-abelian groups,
the Planck scale may correspond to the highest confinement scale. We
comment that the {\it sign} of the coefficient of the  Einstein-Hilbert action,
as inferred from the result of \rf{tv}, comes out right. This is an
important consistency check. Having the right sign means the absence of
perturbative intabilities in the euclidean region, which should clearly be
the case if we are expanding around the correct ground state.

  Having identified the Planck scale with the highest mass scale of the
matter sector,
the next question is whether the DSRT model can provide a consistent
description of the gravitational dynamics at short (\ie shorter than Planckian)
distances. As long as we do not get too close to the lattice scale(s), there
should exist a scaling region where perturbative correlation functions
may depend on dimensionful parameters only through logarithms.

  An interesting possibility, and perhaps the only consistent one,
is that the scaling region is governed by an {\it asymptotically free}
curvature squared~\cite{ftz,csq} effective action.
As shown by Fradkin and Tseytlin, the continuum
curvature squared theory is asymptotically free in all essential
coupling constants~\cite{ftz}. If one demands that the euclidean
action be bounded from below, then the curvature squared theory
in not unitary due to the presence of negative norm states. But the
question of unitarity is really a non-perturbative one~\cite{ftz,csq,igh}.
In order to conclude that negative norm state are indeed present, one has
to analytically continue to Minkowski space and to go on shell.
There are many ways in which non-perturbative effects could invalidate
this troublesome prediction of perturbation theory.

  We conjecture that the perturbative, short distance behaviour of the DSRT
model should be governed a curvature squared theory with a positive definite
euclidean action. This effective short distance theory should arise
from the ``pure gravity'' sector of the model.
By this we mean that, even in the absence of any matter fields,
the improved spacetime measure already defines some effective
gravitational action through eq.~(3.19). We expect this
``pure gravity'' effective action to be of the kind described above.

  One may consider the possibility that, being asymptotically free,
the pure gravity theory could generate a scale dynamically as in QCD.
It is hard to imagine what would be the physics of a model in which
the metric tensor is ``confined''. In any event, in our universe
the gravitational field, like the electromagnetic field, is clearly not
confined in the sense of the gluon field in QCD. We consider this as
an indication that a matter sector must be introduced in order
to generate the Planck scale. An interesting question is whether there is
a relation between the way matter fields should provide the scale which
characterize the gravitational interaction, and the Higgs mechanism which
provides the scale of the Electro-Weak interactions~\cite{igh}.

\vspace{2ex}
\noindent {\bf 3.4~~Beyond General Relativity }
\vspace{1ex}

  In this paper we discussed the possibility that the DSRT model provides
a consistent model of Quantum Gravity. If the considerations we made are
correct, the dynamics of the DSRT model should reduce at sufficiently low
energies to solutions of Einstein's equation coupled self-consistently
to quantum matter.  But under certain circumstances and, in particular, where
General Relativity gives rise to singular solutions,
the physics of the DSRT model should be different and consistent.

  Deviations of the DSRT model from General Relativity may arise from
two sources. The first is the thermodynamical character of the metric tensor
in the DSRT model. Under certain circumstances we anticipate that dissipative
processes will take place. This is expected to occur mainly in the vicinity
of singular points of classical solutions, and so it should be relevant to the
physics of black holes (see below). Under ordinary circumstances, however,
dissipation should be negligible, otherwise one may run into conflict
with observation. A case where modified dynamics may account for
an apparent discrepancy between observation and theory is discussed in \rf{mm}.

  A second source for deviations from General Relativity is the preferred
role of time in the DSRT model. In the domain of validity of General
Relativity, it should be possible to interpret all non-relativistic features
of the DSRT model as non-covariant gauge fixings. But outside the scope of
General Relativity, constraints such as $g^{00}=1$ contain relevant dynamical
information.

  Another feature of a similar nature is the existence of a {\it local
conservation law for the number of space points}. The point density field
$\r(x,t)$ has been introduced earlier. The current density
of space points is  (see eq.~(3.13b))
$$
  J^k = \r\,g^{0k} \,.
\eqno(3.32)
$$
Point number conservation implies the odd-looking conservation equation
$$
  \partial_t \r + \partial_k (\r g^{0k} ) = 0 \,.
\eqno(3.33a)
$$
As we argued earlier, $\r=\sqrt{g}$ dynamically. Therefore,
from the point of view of the low energy curved space, eq.~(3.33a) looks like
a gauge fixing condition too. Although this may be an unconventional
gauge fixing, enforcing it cannot change the physics of General Relativity.

  We can turn around the argument and try to learn under what
circumstances  General Relativity may not work. To this end, we
rewrite the number conservation equation in the following form
$$
  \partial_t(\log\r) + g^{0k}\partial_k(\log\r) + \partial_k  g^{0k}  = 0 \,.
\eqno(3.33b)
$$
the fact that eq.~(3.33b) depends on $\log\r$ suggests that point number
conservation should be relevant, but only if the point density varies by
many orders of magnitude. If this is the case for some physical process,
then General Relativity might not account accurately for
that process.

  Again, in most astronomical systems this is in not the case.
Whenever one has a stationary gravitational system, it should be possible to
find a coordinate system where both $g^{00}=1$ and $g^{0k}=0$. Under these
circumstances, point number conservation holds trivially, and so the
coordinate system may coincide with the preferred coordinate system of the
DSRT lattice.  Other coordinate systems exist in which
$g^{0k}$ does not vanish. As long as both $g^{00}=1$ and eq.~(3.33) hold,
the metric is in principle attainable in the DSRT spacetime measure,
and so these coordinate systems could coincide with the DSRT frame too.

  The above examples show how important it is to identify the
preferred frame of the DSRT lattice with some coordinate system suggested
by the physical process in question. An important case where such an
identification can be made, is if we consider the evolution of entire
universe. It is natural to identify the cosmological time coordinate defined
by the expansion of the universe and the rest frame of the microwave
background radiation,  with the fundamental time coordinate of the DSRT
lattice. Although at this stage we do not have a clear understanding of how
the DSRT model can account naturally for the expansion
of the universe, we hope that this issue will be clarified in the future.
(One can of course enforce the expansion by hand, if one imposes
appropriate initial conditions).

  We conclude with a few comments on the subject of black hole formation and
evaporation. This subject has attracted a lot of attention recently~[17-21].
In the DSRT preferred frame, the components of the metric tensor carry
genuine information about the microscopic properties of spacetime.
The difficulty is to determine how the DSRT frame is related to  curved
coordinate systems used to describe the Schwarzschild metric.

  The Schwarzschild and Kruskal coordinates cannot coincide with the DSRT
frame because they do not satisfy $g^{00}=1$. A coordinate system which does
satisfy this condition and also $g^{0k}=0$ is the Novikov coordinate
system~\cite{nv}. However, the Schwarzschild metric has genuine time
dependence, which might show up as non-trivial time dependence of $\r$
and $g^{0k}$ when expressed in the DSRT frame. Thus, the relation between
the DSRT frame and familiar coordinate systems is not clear at the moment.

  Finding how the Schwarzschild metric looks in  the DSRT frame is in fact
of crucial importance. If the metric tensor turns out to be singular only
at the curvature singularity, then this  supports scenarios~\cite{a,b} in
which the evaporation of a black hole leaves behind a massive, infinitely
degenerate stable remnant. We will not enter at this stage the discussion
of whether or not that stable remnant can be interpreted as a new
asymptotic region. On the other hand, if the metric tensor has a
``coordinate singularity'' in the DSRT frame (say, near the horizon) then
this singularity would not be spurious, and this would support scenarios in
which the horizon acts as a physical membrane~\cite{s,t}.

  Let us return for a moment to the space-space components of the metric
tensor and discuss their physical significance in the DSRT frame. The spatial
metric becomes singular if $\trg_{ij}$ diverges in some region of spacetime,
which is the same as having a zero for $\trg^{ij}$. The field $\trg^{ij}$
measures the local link length squared relative to its global average value,
times the number of links per vertex.
The number of links per vertex is a positive integer. Thus, the
vanishing of $\trg^{ij}$ can arise only from the vanishing of the local
link length, which means that the local point density blows up.

  A singularity in $g_{ij}$ therefore implies that a nearby volume of a
tiny radius in the DSRT frame, will support an enormous number of
matter states compared  to the ``normal'' number of states it supports in
the vacuum. This means that the DSRT model can potentially allow for the
existence of massive stable remnants. Whether or not this is the
way that the paradoxes associated to the evaporation of macroscopic black
holes are resolved, is a question that must await a more detailed
investigation.

\vspace{5ex}
\noindent {\bf 4.~~CONCLUSIONS}
\vspace{3ex}

  The interplay between flat space and curved space has always been an
important theme in the literature on General Relativity~\cite{fc}.
Also, there are numerous attempts to quantize the
gravitational interaction which can be categorized under the title of Induced
Gravity~\cite{igh,ig,cg,ggb}. We believe that the DSRT model goes beyond
previous attempts, and that it provides a unique non-perturbative realization
of the above themes.

  The DSRT model is characterized by the embedding of the spacetime
triangulations in a target flat space, and by the uneven treatment of time
and space. Thanks to the special role of time, the DSRT model is {\it unitary}.
In subsect.~3.1 we argued that, on top of the discretization of time into
regular intervals, one has to let $\ba/a\to 0$
in order to ensure the consistency of the analytic continuation to
Minkowski space. As usual, the price that we pay for having a manifestly
unitary theory, is that restoration of Lorentz covariance is not automatic.
Indeed, we anticipate that under those rare circumstances where General
Relativity fails, the DSRT model may show genuine non-relativistic features.

  One could consider an alternative approach in which time and space are
treated on equal footing. This is in fact a simpler and less elaborate
framework. The spacetime points are allowed to move inside the entire
four dimensional volume and are not constrained to time slices,
while the triangulations are made out of four-simplices.
In this approach, which we denote as the spacetime symmetric approach,
one has the full rotational invariance of four dimension
euclidean space, but the theory is not unitary for finite mean lattice
spacing.

  One may hope that, in this approach, unitarity will be recovered in the
continuum limit. A first signal that this may not be the case is provided
by the discussion of subsect.~3.1. There is no way to impose the analog of
the limit $\ba/a\to 0$ in the spacetime symmetric approach, and so
we suspect that the analytic continuation  may not be consistent in this case.
Also, many of the problems of continuum gravitational theories boil down to
lack of unitarity (see subsect.~3.3) and so there is a serious danger that
they will show up in the spacetime symmetric approach.

  A pessimistic point of view is that, regardless of which option is used,
the embedding of triangulations in a target flat space does not provide a
consistent definition of Quantum Gravity, and that the only difference
between the two options is where the failure occurs. We believe that it
is much too early to draw that negative conclusion. But we do have a clear
cut argument that only the DSRT lattice can possibly give rise to a
consistent theory of Quantum Gravity.

  The argument is based on the Weinberg-Witten (WW) theorem~\cite{ww}.
This theorem asserts that if one has an energy-momentum tensor which satisfies
an ordinary conservation equation and which transforms as a true Lorentz
covariant tensor, then the Fock space of that theory cannot contain
massless spin two states. In General Relativity one avoids the WW theorem
because the true energy-momentum tensor is covariantly conserved. In
an asymptotically flat space one can define another energy-momentum tensor
which satisfies an ordinary conservation equation, but this object is really
a pseudo-tensor.

  In the case of the spacetime symmetric approach, one has an energy-momentum
tensor which satisfies all the assumptions of the WW theorem already
for finite lattice spacing. Hence, the analytic continuation to Minkowski
space cannot  contain gravity and be consistent simultaneously. The DSRT
lattice avoids the WW theorem because it is not a Lorentz covariant framework.
If the DSRT lattice provides a consistent theory of Quantum Gravity, then
relativistic covariance at low energies should be considered as a consequence
of the fact that the underlying dynamics reduces
to solutions of the generally covariant Einstein's equation
(albeit in a special non-covariant gauge).

  The DSRT model has very rich physics, and progress in understanding any
aspect of its physical properties may be important in telling whether or not
it contains gravity. We conclude with a list of possible lines of future
investigation.

  One way is to start directly from the infrared end of the model.
In subsect.~2.4 we proved that the zero momentum scalar state has a finite
energy in the limit of vanishing lattice spacing. It may be possible
to generalized this result to localized states by making use
of variational bounds. Sufficiently detailed information about localized
states may also provide a window to investigate the Planck scale physics
of the model.

  Another possibility is to start from the ultraviolet end. One can investigate
two key features in a simplified version of the model.  These are the method of
enforcing general covariance in the continuum limit (subsect.~3.2) and the
conjecture that  short distance  physics is governed by an asymptotically free
curvature squared theory (subsect.~3.3).

  A nice way to investigate these features is to consider a ``pure measure''
model. This model does not contain any matter fields. We can furthermore
start from a spacetime symmetric model. While in this model we anticipate
difficulties with the analytic continuation to Minkowski space, we believe
that its short distance correlation functions in the euclidean region may be
very similar to those of the DSRT model.

  The spacetime symmetric, pure measure model is defined by the partition
function
\beqn{3.34}
  Z'  & = & Z(N,L,a,\d) \NON
      & = & \int_n \cd y  \int_{x^\m} \cd \x \,,
\eeqn
where
$$
  \int_n \cd y \equiv
    L^{-4N}  \prod_{n=1}^N \int_{-L/2}^{L/2} d^4 y_n \,,
$$
$$
  \int_{x^\m} \cd \x \equiv  (a\d)^{-4(L/a)^4}\, (a\d/L)^4
         \prod_{x^\m} \int d^4 \x_{x^\m}\, \cf_\d(\x),
$$
and $\cf_\d(\x)$ is the spacetime symmetric analog of eq.~(3.22). Here $N$
is the number of spacetime points which are embedded in a four-torus of
circumference $L$ in each direction, and $a$ and $\d$ have the same
significance as in subsect.~3.2.

  The pure measure partition function defines a non-trivial effective
gravitational measure through eq.~(3.19). Apart from the infrared and
ultraviolet cutoffs, it depends on two dimensionless parameters $\d$ and
$\ba/a$ where now $\ba=L/N^{1\over 4}$. One combination of these parameters
should be sent to zero in order to recover general coordinate invariance
in the sense of eq.~(3.20). The other linearly independent combination
is a genuine free parameter.

  If our conjecture concerning the relation between the pure (improved) measure
model and a continuum curvature squared theory is correct, then this second
dimensionless combination should be related to a coupling constant of the
continuum theory. There are in fact two distinct asymptotically free continuum
theories~\cite{ftz} which could be relevant. One is the classically scale
invariant Weyl theory~\cite{csq,igh}. The other is a theory which contains the
two linearly independent curvature squared terms.

  The Weyl theory has one coupling constant. An interesting observation
is that the other curvature squared theory also has only one truely independent
coupling constant which governs the deep euclidean region.  Although
perturbatively one has two coupling constants corresponding to the two linearly
independent curvature squared terms, at the fixed point of the RG flow, the
ratio of the two coupling constants approaches a calculable
constant~\cite{ftz}. The ultraviolet behaviour of both curvature squared
theories is therefore governed by a single free parameter. It would be
interesting to  see whether the pure measure model can be related to one of
these continuum theories.

\vspace{3ex}

  I benefited from discussions with Aharon Casher, Shmuel Elitzur,
Ted Jacobson and Adam Schwimmer. I especially thank Ted Jacobson for his
open-mindedness, as well as for his critical comments, when I discussed with
him very early versions of the ideas developed in this paper.


 \vspace{5ex}
 \centerline{\rule{5cm}{.3mm}}

\end{document}